\documentclass[aps,prd,reprint,superscriptaddress,preprintnumbers,longbibliography]{revtex4-1}

\usepackage{amsmath,amssymb,bm,mathrsfs,bbold,dsfont}
\usepackage{xfrac}
\usepackage[dvipsnames]{xcolor}
\usepackage{physics}
\usepackage{slashed}
\usepackage{multirow}
\usepackage{booktabs}

\usepackage{epsfig}
\usepackage{psfrag}
\usepackage{tikz}
\usepackage{tikz-feynman}
\usepackage{float}
\usepackage[caption=false]{subfig}  
\PassOptionsToPackage{caption=false}{subfig}

\usepackage[sort&compress]{natbib}
\usepackage[colorlinks=true,
    linkcolor=red,
    citecolor=blue,
    urlcolor=blue,
    filecolor=blue,
    anchorcolor=blue,
    menucolor=blue,
    linktocpage=true,
    pdfproducer=medialab,
    pdfa=true
]{hyperref}
\usepackage{cleveref}

\usepackage{setspace}
\usepackage{relsize}
\usepackage{enumerate}
\usepackage{comment}
\usepackage[normalem]{ulem}
\usepackage{srcltx}
\usepackage{xcolor}
\usepackage{diagbox}
\usepackage{appendix}

\newcommand{\emphb}[1]{\textbf{\emph{#1}}}
\newcommand{\GeV}{\,\mathrm{GeV}}
\newcommand{\keV}{\,\mathrm{keV}}
\newcommand{\TeV}{\,\mathrm{TeV}}
\newcommand{\vmin}{v_{\rm min}}

\begin{document}

\preprint{MIT-CTP/6119}

\newcommand{\ourtitle}{Sneaky Sneutrino Scattering at LZ}

\title{\ourtitle}

\author{Kevin Langhoff}
\email{langhoff@mit.edu}
\affiliation{Center for Theoretical Physics -- a Leinweber Institute, Massachusetts Institute of Technology, Cambridge, MA 02139, USA}

\author{Huangyu Xiao}
\email{huangyuxiao@fas.harvard.edu}
\affiliation{
Physics Department, Boston University, Boston, MA 02215, USA
}
\affiliation{Department of Physics, Harvard University, Cambridge, MA 02138, USA
}

\begin{abstract}\noindent
We explore a model of sneutrino dark matter in the Minimal R-Symmetric Supersymmetric Standard
Model as a possible interpretation of the recent event at LZ with high reconstructed nuclear recoil energy. This model can account for a small mass splitting between neutral states in a technically natural way due to terms which violate lepton number; importantly, this model does not require sending masses of other SUSY particles to an extremely high energy scale. Mixing with additional singlet states suppresses the interaction strength with standard model particles which allows the sneutrino to sneak around bounds obtained from searches for annihilating dark matter captured in the sun and allows the dark matter to remain near the TeV scale and still potentially account for the LZ results. Typically, sneutrinos produce too much dark matter unless they are very light; we consider a minimal scenario where production occurs through co-annihilation as one specific mechanism where the dark matter mass is predicted to be around $360$ GeV. 
\end{abstract}

\maketitle


\section{Introduction}\label{sec:Introduction}

Recently, the LUX-ZEPLIN (LZ) collaboration reported the observation of a single event with a reconstructed nuclear recoil energy of $E_R = 248 \pm 23({\rm stat}) \pm 23({\rm sys}){~\rm keV}$~\cite{LZ:2026axp}. One exciting possibility is that this event may have arisen from inelastic scattering with dark matter (DM)~\cite{Tucker-Smith:2001myb, Fan:2026kxx}. The simplest one parameter model which could have explained this event, the thermal higgsino, is excluded by searches for high energy neutrinos from annihilating DM captured by the sun~\cite{Pospelov:2026ewn,IceCube:2025fcu} and is disfavored by the absence of events with higher recoil energy~\cite{Rodd:2026tyn,Pospelov:2026ewn,IceCube:2025fcu}.

In this letter, we consider an alternative model where the Minimal R-Symmetric Supersymmetric Standard Model (MRSSM) is extended by two singlet chiral multiplets which mix with the sneutrino; the lowest mass eigenstate of this mixture makes up all of DM~\cite{Kumar:2009sf,Arkani-Hamed:2000oup,Thomas:2007bu}. 
Approximate $R$ symmetry suppresses Majorana-gaugino-mediated processes, including radiative contributions to neutrino masses and direct annihilation into neutrinos, providing a symmetry motivation for the realization of the proposed model. A small mass splitting between lowest-lying neutral mass eigenstates is protected by approximate lepton-number conservation, while the mixing angle with the singlet suppresses the neutral current which would otherwise lead to dangerous solar capture rates for TeV scale dark matter similar to~\cite{Lee:2026jxl}. We consider a supergravity origin for the associated $R$-breaking soft term, connecting this splitting to the neutrino mass generated through the inverse seesaw. 
This dark matter candidate can make up all of dark matter with a mass near $360$ GeV if produced through co-annihilation, but can also be produced at much larger masses if one considers slightly modified cosmologies or couplings to a dark sector~\cite{Langhoff:2026ujr,Feldstein:2013uha,Giudice:2000ex,Lian:2026hpm}.

A sneutrino interpretation within the NMSSM was considered in~\cite{Lian:2026hpm}. Sneutrinos were emphasized as a motivated inelastic dark matter candidate in the original paper on the subject~\cite{Tucker-Smith:2001myb}, and technically natural small mass splittings between sneutrino CP eigenstates were discussed in~\cite{Grossman:1997is,Hall:1997ah}.

The outline of this paper is as follows: Section~\ref{sec:Model} introduces the model. Section~\ref{sec:Relic abundance} describes the co-annihilation production of dark matter. Section~\ref{sec:lz_solar_results} presents the LZ interpretation and experimental constraints. Section~\ref{sec:Conclusion} summarizes the implications.

\section{The R-Symmetric Sneutrino }\label{sec:Model}

The MRSSM was introduced to alleviate flavor changing neutral currents without requiring flavor-blind SUSY breaking~\cite{Hall:1990hq,Fox:2002bu,Kribs:2007ac}. It imposes a continuous $U(1)_R$ symmetry which forbids Majorana gaugino masses, holomorphic trilinear $A$-terms, and the $\mu H_uH_d$ superpotential term. To give the gauginos and higgsinos masses, the model adds chiral adjoints $\mathcal A_i=(S,T,O)$ with $R=0$ for the three gauge groups, and doublets $R_u,R_d$ with $R=2$ and hypercharges opposite to $H_u,H_d$. Resulting gauginos and higgsinos have Dirac masses. In this work, we assume adjoint scalars, $R$-Higgs scalars, Dirac gauginos, and Higgsinos are at the multi-TeV scale and can be integrated out and dark matter phenomonology is approximated by slepton and singlet sectors alone.

\emphb{Singlet Sector.} We introduce two SM singlet chiral multiplets, $N$ and $\bar{N}$, which both have $U(1)_R$ charge $1$ and lepton number $\pm 1$ respectively. For simplicity, we work with one generation. The superpotential terms involving these fields are
\begin{align}
    W\supset Y_\nu H_u L \bar{N} + M_N N\bar{N} + \frac{\mu_X}{2}NN,
\end{align}
where $\mu_X$ is assumed to be the only source of lepton number violation such that small $\mu_X$ is technically natural. There could be a term $\bar{N}\bar{N}$, but this wouldn't affect the rest of the story so we ignore it.

Let us momentarily set $\mu_X = 0$ such that the scalar component of $\bar{N}$ decouples from the other neutral scalar fields which, after electroweak symmetry breaking, have a squared mass matrix in the $(\tilde{\nu}_L,\,\tilde{N})$ basis
\begin{align}
    \mathcal{M}^2 = 
    \begin{pmatrix}
        m_{\tilde{L}}^2 + m_D^2+\frac{1}{2}m_Z^2 \cos(2\beta) & m_D M_N\\
        m_D M_N & m_{\tilde{N}}^2 + M_N^2
    \end{pmatrix}
\end{align}
where $m_{\tilde{L}}$ and $m_{\tilde N}$ are soft masses and $m_D = Y_\nu v \sin \beta/\sqrt{2}$ (with $v = 246\,{\rm GeV}$). The term proportional to $m_Z^2$ arises from $D$-terms; $D$-terms can be decreased relative to this value depending on the chiral adjoint mass spectrum, but our conclusions are relatively insensitive to such modifications, which we ignore.

The light mass eigenstate, with mass $m_\chi$, is given by 
\begin{align}
    \phi = \cos \theta\, \tilde{N} - \sin \theta\, \tilde{\nu}_L,\quad \tan 2\theta = \frac{2\,m_D\,M_N}{\mathcal{M}_{11}^2 - \mathcal{M}_{22}^2},
\end{align}
where $\tilde{\nu}_L$ is the active component from the doublet $\tilde{L}$; $\phi$ therefore realizes the mixing suppression with the $Z$ boson described in~\cite{Lee:2026jxl}. In the following we will be interested in suppressing this interaction by $\theta \sim \mathcal{O}(0.1)$.

The orthogonal mass eigenstate
$\Phi=\sin\theta\,\tilde N+\cos\theta\,\tilde\nu_L$
has mass $m_\Phi=m_\chi+\Delta$, with 
\begin{align} \label{eq:Delta}
   \Delta \approx  \frac{m_D M_N}{m_\chi \sin (2\theta)}
\end{align}
For co-annihilation, we require $\Delta \sim \mathcal{O}({\rm few\, GeV})$ which requires $\mathcal{O}(1\%)$ tuning in soft masses. However, since this story allows SUSY to remain near the TeV scale, as opposed to the higgsino story where gauginos are around $10\, \rm PeV$ in mass, this is not too bad. 

For $\Delta \ll m_\chi$, the charged slepton mass, $m_{\tilde{\ell}_L}$, obtains a mass splitting from electroweak $D$ terms
\begin{align} \label{eq:Delta_charged}
   m_{\tilde{\ell}_L}^2   \approx m_\chi^2+ m_W^2 |\cos (2\beta)|.
\end{align}
For example, if $m_\chi\approx 300\GeV$ this gives $m_{\tilde{\ell}_L} - m_\chi \approx 10\GeV$; which would make them an interesting target for compressed SUSY searches at the LHC~\cite{ATLAS:2019lng} or electroweak precision tests at the FCC-ee~\cite{Knapen:2024bxw}. Potential modifications to $D$-terms in the MRSSM could decrease this mass splitting further.

Supergravity terms which break the $U(1)_R$ global symmetry arise proportional to the gravitino mass~\cite{Hall:1983iz}. A lepton violating contribution proportional to the lepton number spurion $\mu_X$ arises of the form
\begin{align}
    V_{\slashed{R}}\supset \frac{b_N}{2}\tilde{N}^2 + {\rm h.c.},\quad b_N \sim m_{3/2}\,\mu_X.
\end{align}
We emphasize the importance of lepton number in this model; if we instead considered the possibility of a dirac gaugino as an inelastic dark matter candidate, we would expect a gravitationally generated majorana mass of order $m_{3/2}$ which either ruin compatibility with small mass splitting or predict a gravitino LSP. This generates a mass splitting in the real and imaginary part of $\phi$. Define these mass eigenstate components as $\chi_1$ and $\chi_2$, with masses $m_\chi \mp \delta/2$ respectively, where 
\begin{align}
     \delta=\frac{|b_N|}{m_\chi}\cos^2\theta .
\end{align}
Importantly, this splitting can be parametrically smaller than the mass of dark matter due to the inherited technical naturalness from $\mu_X$.

\emphb{Off-Diagonal Neutral Current.}  In the interaction basis, the active sneutrino has neutral current interaction
\begin{align}
    \mathcal{L}\supset \frac{g_Z}{2}Z^\mu J_\mu,\quad J_\mu = \Big(i\tilde{\nu}_L^*\partial_\mu \tilde{\nu}_L + {\rm h.c.}\Big),
\end{align}
where $g_Z = g/\cos \theta_W$. Using $\tilde{\nu}_L = -\sin \theta \,\phi + \cos \theta \Phi$ and $\phi = (\chi_1 + i \chi_2)/\sqrt{2}$ gives the following off diagonal neutral current interaction with the mass eigenstates:
\begin{align}
     \mathcal{L}\supset \frac{g_Z \sin^2 \theta}{2}Z^\mu J_\mu,\quad J_\mu = \Big(\chi_2\partial_\mu \chi_1 - \chi_1\partial_\mu \chi_2\Big).
\end{align}
Tree-level $Z$-mediated scattering is therefore inelastic, $\chi_1+A\to\chi_2+A$, with an amplitude suppressed by $\sin^2\theta$.

\emphb{Neutrino Masses.} Before interpreting this theory as dark matter, let us first connect this story to neutrino masses. We restrict to one generation which limits our predictability for flavor physics, but we obtain an interesting relation with the neutrino mass scale.

In the basis $(\nu,\, N,\, \bar{N})$, the mass matrix is 
\begin{align}
    \mathcal{M}_\nu = 
    \begin{pmatrix}
    0 & 0 & m_D\\
    0 & \mu_X & M_N\\
    m_D & M_N & 0
    \end{pmatrix}\implies m_\nu = \frac{m_D^2\mu_X}{M_N^2}.
\end{align}
Trading the superpotential parameters for those describing late time dark matter phenomenology gives
\begin{align}
     m_\nu &\approx  \frac{4\,\delta\,\Delta^2\,m_\chi^3 \sin^2\theta}{m_{3/2}\,M_N^4}\\
    &\approx 0.01\,{\rm eV}\,\left(\frac{3\TeV}{m_{3/2}}\right)
      \left(\frac{300\GeV}{m_\chi}\right)\left(\frac{m_\chi}{M_N}\right)^{4}\notag\\
    &\qquad\quad ~~ \times \left(\frac{\delta}{250\,{\rm keV}}\right)
      \left(\frac{\Delta}{\GeV}\right)^{2}\left(\frac{\theta}{0.1}\right)^{2}.
\end{align}
Therefore, choosing $m_\chi \sim M_N$ (motivated by naturalness, since $M_N\gg m_\chi$ needs careful cancellation with $m_{\tilde{N}}$ for small $m_\chi$) and $m_{3/2}\sim \mathcal{O}(\TeV)$ gives $(m_\chi,\,\delta,\,\Delta,\, \theta)$ consistent with the correct DM relic abundance via co-annihilation, the LZ data, avoiding solar capture bounds, and the observed SM neutrino mass scale.

\section{Relic abundance from co-annihilation}\label{sec:Relic abundance}
Now we compute the relic abundance from a coannihilation scenario, provided that the mass splitting between active sneutrinos and $\chi$ is only a few GeV. In such scenarios, the annihilation processes are not suppressed by a high power of ${\rm sin}\theta$. Since the conversion rate between active and sterile sneutrinos is much larger than the Hubble rate, the relevant annihilation processes are dominated by those unsuppressed ones involving active sneutrinos and sleptons. In Appendix.~\ref{app:eft}, we discuss an IR model that matches the supersymmetric theory.
The annihilation of many supersymmetric particles can be described by an effective cross-section~\cite{Edsjo:1997bg}
\begin{equation}
    \frac{dn}{dt}=-3H n-\langle \sigma_{\rm eff} v\rangle (n^2-n_{\rm eq}^2),
\end{equation}
with the effective cross-section given by summing over different annihilation channels: 
\begin{equation}
\langle \sigma_{\mathrm{eff}} v \rangle
= \sum_{ij} \langle \sigma_{ij} v_{ij} \rangle
  \frac{n_i^{\mathrm{eq}}}{n^{\mathrm{eq}}}
  \frac{n_j^{\mathrm{eq}}}{n^{\mathrm{eq}}} \, .
\end{equation}
The annihilation cross-section can be formally written as
\begin{equation}
    \sigma_{ij }=\sum_X \sigma(\chi_i\chi_j\rightarrow X),
\end{equation}
where $X$ represents different final states of the annihilation process and $\chi_i$ are different annihilating particles. The relevant channels are summarized in Table.~\ref{tab:coannihilation_channels}.

\begin{table}[t]
    \centering
    \renewcommand{\arraystretch}{1.15}
    \begin{tabular}{ll}
        \hline\hline
        Initial state & Final states   \\
        \midrule
        $\Phi\Phi^*$
        & $W^+W^-,\ ZZ,\ hh,\ Zh,$  $f\bar f$ \\
        \addlinespace
        $\Phi\tilde\ell_L^+, \Phi^*\tilde\ell_L^-$ 
        & $W^\pm\gamma,\ W^\pm Z,\ W^\pm h$  $f\bar f'$ \\
        \addlinespace
        $\tilde\ell_L^-\tilde\ell_L^+$
        & $\gamma\gamma,\ \gamma Z,\ ZZ,\ W^+W^-$ \\
        \hline\hline
    \end{tabular}
    \caption{Main coannihilation channels at leading
    order. The
    $\Phi\Phi^*\to W^+W^-,ZZ,hh$ channels dominate the benchmark when
    kinematically accessible; the remaining rows list important
    subleading channels. Here $f$ denotes an SM fermion, while
    $f\bar f'$ denotes an allowed charged-current fermion pair with
    the same total charge as the initial state. }
    \label{tab:coannihilation_channels}
\end{table}

\begin{figure}[t!]
    \centering
    \includegraphics[width=\columnwidth]{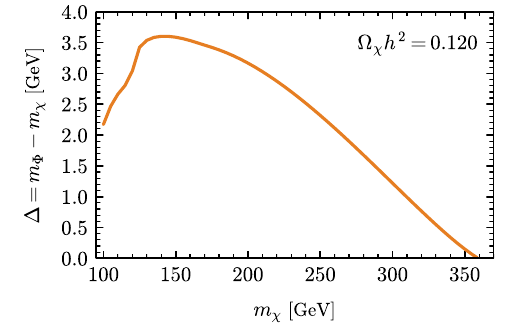}
    \caption{Parameters producing the observed relic abundance in the $(m_\chi,\Delta)$ plane,
    where $\Delta=m_\Phi-m_\chi$.
    The calculation includes gauge and Higgs channels at leading
    order.}
    \label{fig:relic_contour}
\end{figure}

Since we are always in the regime $\delta\ll T_f \sim m_\chi/25$, the inelastic mass splitting does not affect the relic abundance. When ${\rm sin} \theta$ is small, it also does not change the mass spectrum significantly. Therefore, the relic abundance is dominantly set by $m_{\chi}, \Delta$. We compute the annihilation cross section and the relic abundance using MicrOMEGAs \cite{Belanger:2001fz}. The contour of parameters leading to the correct dark matter abundance is shown in Fig.~\ref{fig:relic_contour}. To successfully produce the correct relic abundance, we need $m_\chi<360$ GeV. The thermal target of a scalar electroweak doublet is around 540 GeV \cite{Cirelli:2005uq}. For this coannihilation scenario, since there are 4 degrees of freedom annihilating out of 6 total degrees of freedom, it is naturally expected that the thermal target is $2/3$ of the usual scalar doublet. At $m_\chi>360$ GeV, coannihilation is not sufficient to annihilate them away, and the $\chi$ abundance is overproduced. However, nonstandard thermal history that injects additional entropy can easily dilute the abundance and reproduce the correct relic density at higher masses~\cite{Langhoff:2026ujr,Feldstein:2013uha,Giudice:2000ex,Lian:2026hpm}.

\begin{figure*}[t!]
\centering
\includegraphics[width=\textwidth]{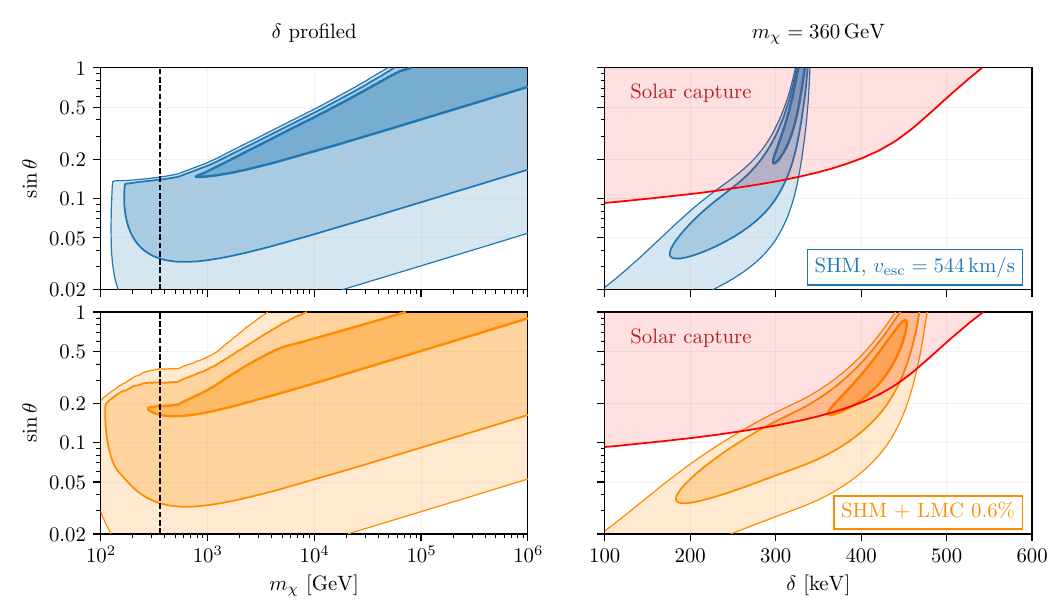}
\caption{LZ interpretation and solar-neutrino bounds for the two halo prescriptions. The top row shows the standard halo model with $v_{\rm esc} = 544$ km/s while the second row shows the result for a DM halo with contributions from the large Magellanic Cloud.  First column: the likelihood is profiled over $\delta$ subject to the solar bound. Second column: LZ contours at $m_\chi=360\GeV$, indicated by the dashed lines above. Red denotes the bounds obtained from solar capture. Contours are labeled from dark to light shading as $\Delta\chi^2<2.30$, $6.18$, and $11.83$, using the unrestricted three-parameter maximum separately for each halo; these are heuristic likelihood contours. }
\label{fig:lz_solar}
\end{figure*}

\section{Fit to LZ and Solar Capture}
\label{sec:lz_solar_results}
We now consider how the above model fits to late time experimental observations; namely, the observed high energy recoil event at LZ and the absence of other events at different nuclear recoil energies~\cite{LZ:2026axp} as well as the absence of excess high energy neutrino flux from the sun observed at IceCube~\cite{IceCube:2025fcu}. The inelastic scattering signal and solar capture rate depend on $m_\chi$, $\delta$, and $\sin\theta$. The high energy neutrino flux from the sun also depends on the annihilation rate, which we discuss below.

\emphb{Fit to LZ Data.} First we fit the model to the LZ data, which in this case is given by the single event with $E_R = 248 \pm 23({\rm stat}) \pm 23({\rm sys}){~\rm keV}$, no other events in the region of interest ($E_R < 270\keV$), and no events in the high energy sideband ($350\keV\lesssim E_R\lesssim 675\keV$).

The differential nuclear recoil rate is nearly identical to the case of the higgsino as the scalar current contributes identically in the non-relativistic limit. There is also the additional suppression by $\sin^4\theta$ such that
\begin{align} \label{eq:rate}
    \frac{dR}{dE_R} = \frac{\rho_\chi \sigma^0_{A}Q_W^2}{2m_\chi \,\mu_A^2}\,F^2(E_R)\, \eta(\vmin(E_R)),
\end{align}
where $\rho_\chi$ is the local DM density (taken to be $0.3\,\GeV/{\rm cm}^3$), $\sigma_A^0 = G_F^2\mu_A^2 \sin^4\theta/2\pi$ is the $Z$-exchange cross section at zero momentum exchange~\cite{Essig:2007az,Rodd:2026tyn}, $\mu_A$ is the reduced mass, $F$ the Helm form factor~\cite{Helm:1956zz,Lewin:1995rx}, $Q_W = N - (1-4s_W^2)Z$, and $\eta(\vmin) = \int_{\vmin} d^3v\, f(\bm v)/v$ in the laboratory frame, averaged over a year~\cite{McCabe:2010zh}. Importantly, the minimal relative velocity to cause a nuclear recoil with energy $E_R$ is 
\begin{align} \label{eq:vmin}
\vmin(E_R) &= \sqrt{\frac{\delta}{2\mu_A}}\left(\sqrt{\frac{E_R}{E_R^*}} + \sqrt{\frac{E_R^*}{E_R}}\right),
\end{align}
with $E_R^* = (\mu_A/m_A)\,\delta$. Boltzmann suppression causes $\eta(\vmin)$ to be a very steeply falling function with increasing $\vmin$ such that kinematics alone strongly enhances collisions for nuclear recoil events near $E_R^*$; the form factor, for the recoil energies of interest, pulls the actual peak of the recoil distribution to slightly lower values.

Since inelastic scattering probes the high velocity tail of the DM halo velocity distribution, we consider two benchmark halo models. The Standard Halo Model (SHM) given by $f_{\rm SHM}(\mathbf v)\propto e^{-v^2/v_0^2}\Theta(v_{\rm esc}-v)$ in the galactic frame with $v_0 = 238$ km/s and $v_{\rm esc} = 544$ km/s~\cite{Baxter:2021pqo}. We also consider $f=(1-w)f_{\rm SHM}+wf_{\rm LMC}$ with $w=0.006$ following~\cite{Rodd:2026tyn}, where $f_{\rm LMC}$ is the high-speed population associated with the Large Magellanic Cloud~\cite{Besla:2019xbx,OHare:2026nqi}.

Following Ref.~\cite{Langhoff:2026ujr}, we use a likelihood
\begin{align}
    -\log\mathcal L = N_{\rm ROI}+N_{\rm SB}-\log p(248\keV),
\end{align}
where $p$ is the expected, unnormalized signal-event density after detector response and acceptance. We fix the detector response and quote $\Delta\chi^2=-2\log(\mathcal \mathcal L/L_{\max})$ relative to the unrestricted maximum over $(m_\chi,\delta,\sin\theta)$ separately for each halo, before imposing solar or relic constraints. Figure~\ref{fig:lz_solar} shows contours at $2.30$, $6.18$, and $11.83$ as heuristic measures of relative fit quality. 

\emphb{Solar Capture Bounds.} The same inelastic interaction can capture sneutrinos in the Sun, where gravitational acceleration allows up-scattering on solar nuclei even when terrestrial scattering samples only the fastest halo particles~\cite{Nussinov:2009ft,Menon:2009qj,Pospelov:2026ewn}. The capture rate inherits the mixing suppression as
\begin{align}
C_\odot(m_\chi,\delta,\theta)=\sin^4\theta\,C_0(m_\chi,\delta),
\end{align}
where $C_0$ is the capture rate without mixing suppression. Repeated inelastic collisions cool the captured particles until up-scattering becomes inaccessible. Further cooling then depends
on elastic scattering.

For sneutrinos this cooling proceeds through a tree-level Higgs interaction, $\mathcal{L}\supset- \frac{1}{2}g_{h\chi\chi}h\chi_1^2$ with $g_{h\chi\chi} \approx \frac{1}{2}g_Z m_Z \sin^2 \theta$. The sneutrino has spin-independent nucleon cross section
\begin{align}
    \sigma_n^{\rm SI}
    = \frac{\mu_{\chi n}^2}{4\pi m_\chi^2}
      \left(
      \frac{f_n\, m_n\, g_{h\chi\chi}}{v \,m_h^2}
      \right)^2 ,
\end{align}
where $\mu_{\chi n}$ is the reduced mass and $f_n$ is the nucleon form factor. Unlike the accidentally suppressed, loop-induced spin-independent amplitude of the pure higgsino, this interaction is present at tree level. For the benchmarks considered, elastic scattering brings the captured sneutrinos into kinetic equilibrium with the solar plasma.

For a DM population in kinetic equilibrium with the solar plasma, the annihilation rate is
\begin{align}\label{eq:solar_annihilation}
    \Gamma_A
    &=\frac{C_\odot}{2}
      \tanh^2\!\left(\frac{t_\odot}{\tau_{\rm eq}}\right),
    &
    \tau_{\rm eq}
    &=\left(
    \frac{V_{\rm eff}}
    {C_\odot\langle\sigma v\rangle}
    \right)^{1/2}.
\end{align}
Here $V_{\rm eff}$ is the effective annihilation volume, $\langle\sigma v\rangle$ is the annihilation cross section averaged over that population, and $\tau_{\rm eq}$ is the capture-annihilation equilibration time~\cite{Nussinov:2009ft,Pospelov:2026ewn}. Kinetic equilibrium fixes $V_{\rm eff}$, but equilibrium between the solar capture and annihilation processes additionally requires $\tau_{\rm eq}\ll t_\odot$. In our leading electroweak approximation, $C_\odot,\langle\sigma v\rangle\propto\sin^4\theta$, so that $\tau_{\rm eq}\propto\sin^{-4}\theta$ at fixed masses and thermal distribution. Mixing can therefore suppress the solar-neutrino signal even when elastic scattering efficiently brings the captured population into kinetic equilibrium. 

We compare with the IceCube $W^+W^-$ annihilation-rate limit \cite{IceCube:2025fcu}. The resulting bounds are shown in Fig.~\ref{fig:lz_solar}. Above $10\TeV$, we hold the annihilation-rate limit fixed following Ref.~\cite{Langhoff:2026ujr}.

\emphb{Elastic Scattering.} Although the $Z$ current is inelastic, Higgs exchange also mediates elastic nuclear scattering. The cross section quoted above lies below the published LZ elastic-scattering limits~\cite{LZ:2024elastic}, while still allowing cooling of the captured solar population.

\emphb{Collider Bounds.} Sneutrino dark matter could potentially be observed at the LHC through the decay of charged sleptons. Their compressed spectrum gives soft visible decay products. Here the charged partner instead decays through an off-shell $W$ which has relatively weak bounds for $m_{\tilde{\ell}} - m_{\chi} < m_W$~\cite{Carpenter:2020sneutrino} and can be ignored for the masses of interest.

\emphb{Indirect Detection.} Partners participating in co-annihilation are absent today, leaving a mixing suppressed annihilation rate. At $m_\chi=360\GeV$ and $\sin\theta=0.1$, this is approximately $10^{-29}\,\mathrm{cm^3\,s^{-1}}$, well below continuum gamma-ray limits for electroweak final states from Fermi-LAT and H.E.S.S.~\cite{FermiLAT:2015dwarfs,HESS:2022IGS}. Energy injection into the CMB is also small with $\langle\sigma v\rangle_0/m_\chi\lesssim10^{-31}\,\mathrm{cm^3\,s^{-1}\,GeV^{-1}}$; three orders of magnitude below the Planck bound~\cite{Planck:2018cosmo}. These comparisons use perturbative annihilations, although the conclusion remains unless Sommerfeld enhancement is extremely large.

\begin{table}[t]
\centering
\renewcommand{\arraystretch}{1.15}
\setlength{\tabcolsep}{2pt}
\begin{tabular}{
    l@{\hspace{10pt}\vrule width 1pt\hspace{10pt}}ccc
    @{\hspace{10pt}\vrule width 1pt\hspace{10pt}}
    ccc
}
\hline\hline
& \multicolumn{3}{c@{\hspace{10pt}\vrule width 1pt\hspace{10pt}}
  }{SHM}
& \multicolumn{3}{c}{SHM + LMC $0.6\%$} \\
$m_\chi$/TeV
& 0.36 & 1 & 10
& 0.36 & 1 & 10 \\
\hline
\hline
$\delta$/keV
& 282 & 319 & 360 & 376 & 412 & 442 \\
$\sin\theta$
& 0.129 & 0.158 & 0.401 & 0.178 & 0.219 & 0.439 \\
$N_{\rm ROI}$
& 0.753 & 0.652 & 0.594 & 0.602 & 0.557 & 0.605 \\
$N_{\rm SB}$
& 0.008 & 0.082 & 0.379 & 0.066 & 0.184 & 0.395 \\
$E_{\rm peak}$/keV
& 165 & 178 & 198 & 175 & 184 & 192 \\
$\Delta\chi^2$
& 2.92 & 2.17 & 1.66 & 2.18 & 1.94 & 1.84 \\
$\Gamma_A/\Gamma_{\rm lim}$
& 1.00 & 1.00 & 1.00 & 1.00 & 1.00 & 0.09 \\
\hline\hline
\end{tabular}
\caption{
Best fits at fixed sneutrino masses, maximizing the LZ likelihood over $\delta$ and $\sin\theta$ subject to solar capture bounds. Halo prescriptions use $v_{\rm esc}=544\,{\rm km/s}$. $E_{\rm peak}$ denotes maxima of the nuclear recoil distributions.
}
\label{tab:sneutrino_fits}
\end{table}

\section{Conclusion}\label{sec:Conclusion}

While only a single event, the high nuclear recoil event seen at LZ has the very exciting potential to be our first observation of dark matter. If this is actually the case, it hints at dark matter which may differ from our simplest expectation and warrants a thorough exploration of potential explanations (for example,~\cite{Langhoff:2026ujr,
Su:2026rwz,Lou:2026idn,Nomura:2026qyq,Yamashita:2026ump,%
Chattopadhyay:2026ryw,Smirnov:2026aqk,Du:2026guj,McCabe:2026crm,%
Jeesun:2026vzo,Unwin:2026rdp,Dent:2026bji,deLima:2026shq,%
Gu:2026vto,Baer:2026fpy,Lee:2026wof,Wang:2026ytg,%
Yang:2026wpb,Kotlarski:2026pep,Liang:2026coz,Das:2026uyy,%
Alhazmi:2026efz,Okada:2026eol,Ahmed:2026qjg,Du:2026lpa,%
Bandyopadhyay:2026gjw,Kannike:2026qyl,Borah:2026zwf,Bisal:2026khf,%
Cheung:2026byg,Yuan:2026djt,Elahi:2026vlm,Zhu:2026dag,%
Aghaie:2026vsu,Asadi:2026iot,Lee:2026xxh,Lee:2026jxl,%
Khan:2026nwp,He:2026hqz,Fan:2026hzw,Chattaraj:2026fxn}).

In this paper, we have considered the possibility that dark matter could be a sneutrino from the MRSSM extended by two singlet chiral multiplets. This allows a technically natural small mass splitting consistent with the nuclear recoil energy seen at LZ and also allows for supression of the interaction with the SM by mixing with the SM singlet chiral multiplets to alleviate solar capture bounds. This gives a simple UV complete theory compatible with observation where all new particles are allowed to have masses remaining near the TeV scale.

Coannihilation provides one predictive realization at a few hundred GeV, with parameter space compatible with a LZ interpretation and solar bounds. The fit with LZ also admits a few-TeV sneutrino (see Tab.~\ref{tab:sneutrino_fits}). Such massive sneutrino DM can account for the observed relic abundance if there is entropy dilution (e.g.~\cite{Langhoff:2026ujr,Feldstein:2013uha}) or if additional MSSM or dark-sector interactions are considered (e.g.~\cite{Arina:2008bb,Lian:2026hpm}). The simplicity of TeV scale sneutrino DM warrants further investigation in this direction.

\begin{acknowledgments}
\noindent 
We thank Patrick Fox, Matthew Reece, Neal Weiner and Lisa Randall for helpful discussions and valuable comments on the draft.
This material is based upon work supported by the U.S. Department of Energy, Office of Science, Office of High Energy Physics of U.S. Department of Energy under grant Contract Number  DE-SC0012567."(High Energy Theory research) and Simons Foundation Investigation Awards 929255 and 929241. HX is supported by the U.S. Department of Energy under grant DE-SC0026297.
\end{acknowledgments}

\appendix
\section{Effective Lagrangian}\label{app:eft}
In our model, an approximately R-symmetric supersymmetric theory motivates a light mixed
sterile--active sneutrino, with small R-breaking soft terms providing an
inelastic splitting. We can describe the effective model by introducing an EW singlet and doublet:
\begin{equation}
    \tilde{N}\sim(\mathbf1,0),\qquad \tilde{L}=\begin{pmatrix}
     \tilde{\nu}_L \\ \ell_L^- 
    \end{pmatrix}\sim(\mathbf2,-\tfrac12).
\end{equation}
The gauge and Higgs interactions of these mixed scalars are given by 
\begin{align}
       \mathcal{L}\supset & |\partial_\mu \tilde{N}|^2+|\mathcal{D}\tilde{L}|^2+ m_{\tilde{N}}^2 |\tilde{N}|^2+ m_{\tilde{L}}^2 \tilde{L}^2+ \\ &\left( a \tilde{N} \tilde{L}\cdot H+\frac{b_N}{2}\tilde{N}^2+\rm h.c.\right)
\end{align}
 
The singlet and doublet scalars are mixed through $ a \tilde{N} \tilde{L}\cdot H$, generating a mass splitting. The lowest mass eigenstate ($\phi$) is mostly sterile and the active one ($\Phi$) is heavier:
\begin{equation}
    \phi = c_\theta \tilde{N}-s_\theta \tilde{\nu}_L, \qquad \Phi= s_\theta \tilde{N} +c_\theta \tilde{\nu}_L,
\end{equation}
with the mixing angle given by
\begin{equation}
    {\rm tan}\,2\theta = \frac{\sqrt{2} a v}{m_{\tilde{L}}^2-m_{\tilde{N}}^2}. 
\end{equation}
This matches the UV theory with the parametrization $m_D m_N=av/\sqrt{2}$.
We further decompose the complex scalar as
\begin{equation}
    \phi = \frac{1}{\sqrt{2}}(\chi_1+i\chi_2), \qquad \Phi= \frac{1}{\sqrt{2}}(\chi_3+ i \chi_4). 
\end{equation}
The gauge interactions are fixed by the doublet representation as well as its mixing with the singlet. Choosing basis with $\varphi=(\phi,\Phi, \ell_L^{-} )$, the gauge interactions can be written as
\begin{align}
    \mathcal{L}_{\rm gauge} & = |(\partial_\mu-i\mathcal{G}_\mu)\varphi|^2 \\
    &=|\partial_\mu \varphi|^2+\varphi^\dagger\mathcal{G}_\mu\mathcal{G}^\mu\varphi+i\varphi^\dagger\mathcal{G}_\mu \partial_\mu\varphi-i(\partial^\mu\varphi^\dagger)\mathcal{G}_\mu\varphi,
\end{align}
with the gauge connection
\begin{equation}
    \mathcal{G}_\mu = \begin{pmatrix}
       \frac{g_Z}{2}s_\theta^2 Z_\mu& -\frac{g_Z}{2}s_\theta c_\theta Z_\mu & -g s_\theta W_\mu^{+}/\sqrt{2}\\
       -\frac{g_Z}{2}s_\theta c_\theta Z_\mu & \frac{g_Z}{2} c_\theta^2 Z_\mu & g c_\theta W_\mu^+/\sqrt{2} \\
       -gs_\theta W_\mu^{-}/\sqrt{2} & g c_\theta W_\mu^{-}/\sqrt{2} & -eA_\mu +g_C Z_\mu
    \end{pmatrix},
\end{equation}
Where $g_C=g_Z(-1/2+s_W^2)$ with $g$ being the $SU(2)$ gauge coupling.
The gauge interactions of $\phi,\Phi$ are determined by their overlap with the scalar doublet. In particular, the light-current is inelastic scattering through exchanging $Z$ bosons
\begin{equation}
    \mathcal{L}_Z \supset \frac{gs_\theta^2}{2c_W} Z_\mu (\chi_2 \partial^\mu \chi_1-\chi_1\partial^\mu\chi_2).
\end{equation}

The mixing term $a \tilde{N} \tilde{L}\cdot H$ also generates Higgs couplings
\begin{equation}
    \mathcal{L}_h \supset -\frac{\delta g_{h\phi\phi}}{2}h(\chi_1^2+\chi_2^2), \qquad \delta g_{h\phi\phi}= -\sqrt{2} a\, s_\theta c_\theta.
\end{equation}
Using the mass-mixing relation, this contribution becomes
\begin{equation}
    \delta g_{h\phi\phi}
    =-\frac{m_\Phi^2-m_\chi^2}{2v}\sin^2(2\theta)
    \simeq -\frac{4m_\chi\Delta}{v}
    s_\theta^2 c_\theta^2.
\end{equation}
Thus, at fixed mass spectrum, the Higgs-mediated
annihilation of the light sneutrino is suppressed by
$\sin^4\theta$ and is subleading to the unsuppressed
coannihilation processes. Additionally, this mixing-induced Higgs coupling is usually smaller than the expected $D$ term coupling in the SUSY model that gives $g_{h\phi\phi}\approx -\frac{1}{2}g_Zm_Z\, {\rm sin}\theta^2$.

The predominantly active sneutrino and charged slepton
also have Higgs couplings that are not suppressed by the
active--sterile mixing angle. Assuming ordinary supersymmetric $D$-term quartics and
neglecting small Yukawa and mixing corrections, these
interactions are
\begin{equation}
    \mathcal{L}_h \supset
    -\frac{g_Z^2\cos(2\beta)}{4}\left(vh+\frac{h^2}{2}\right)
    \left(|\Phi|^2
    +(1-2c_W^2)|\tilde{\ell}_L^-|^2\right).
\end{equation}
We take $\cos(2\beta)\simeq-1$ for our numerical benchmark.
These interactions survive in the matched IR theory and
introduce no additional free parameters once this matching
limit is specified.  
We include these contributions in the relic-abundance
calculation. For the R-Higgs sector in the full theory, we assume $R_{u,d}$ and the Higgsino to be heavy such that they do not contribute to the relic density calculation.

When the mass gap between the active and sterile sneutrino ($\Delta$) is small, co-annihilation can produce the correct relic abundance for the nearly sterile sneutrino dark matter.
The relevant derivative vertices are approximately given by
\begin{align}
    \mathcal{L}_3= &i \frac{g_Z}{2}Z_\mu (\Phi^*\partial^\mu \Phi -\Phi\partial^\mu\Phi^*)\\
   & +i(-eA_\mu +g_C Z_\mu)(\tilde{\ell}^+_L\partial^\mu\tilde{\ell}^-_L-\tilde{\ell}^-_L\partial^\mu\tilde{\ell}^+_L)\\
   &+\left[\frac{ig}{\sqrt{2}}W_\mu^+(\Phi^*\partial^\mu\tilde{\ell}^-_L -\tilde{\ell}^-_L\partial^\mu\Phi^*)+\rm h.c.\right],
\end{align}
and the contact vertices are
\begin{align}
    \mathcal{L}_4 =&\left(\frac{g_Z^2}{4}Z_\mu Z^\mu + \frac{g^2}{2}W_\mu^+W_\mu^- \right)|\Phi|^2\\
    &+\Bigl(e^2 A_\mu A^\mu -2eg_CA_\mu Z^\mu +g_C^2Z_\mu Z^\mu\notag\\
    &\hspace{1.5em}+\frac{g^2}{2}W_\mu^+W^{\mu-}\Bigr)|\tilde{\ell}_L|^2\\
    &+\left(\frac{g}{\sqrt{2}}W_\mu^+[{g_Z}s_W^2Z^\mu-eA^\mu]\Phi^*\tilde{\ell}_L^-+\rm h.c.\right),
\end{align}
where we approximated $c_\theta\approx 1$ and $s_\theta\approx0$ to get rid of the suppressed channels and used $g_C=g_Z(-1/2+s_W^2)$.
\bibliography{bibliography}

\end{document}